\documentclass{aa}
\usepackage{graphicx, hyperref}
\usepackage{txfonts}
\usepackage[noabbrev]{cleveref}
\def\dustp{DUSTGRAIN-\textit{pathfinder} }
\usepackage{mathrsfs}
\usepackage[normalem]{ulem}
\begin{document}
\title{Breaking the cosmological invariance of the dark-matter halo shape as a new probe of modified gravity}
\titlerunning{Breaking of the halos shape as a new probe of MG}
\author{R\'emy Koskas \and Jean-Michel Alimi}
\institute{Laboratoire Univers et Théories (LUTH), CNRS, Observatoire de Paris, PSL Research University\\
5 place Jules Janssen, 92195 Meudon, France\\
\email{remy.koskas@obspm.fr}; \email{jean-michel.alimi@obspm.fr}}
\date{Received X / Accepted Y}
\abstract{In a recent paper, Alimi \& Koskas highlighted in $w$CDM models derived from general relativity (GR) (with Dark Energy Universe numerical simulation data), a cosmological invariance of the distribution of dark-matter (DM) halo shapes when expressed in terms of the nonlinear fluctuations of the cosmic matter field. This paper shows that this invariance persists when tested on numerical simulations performed with a different N-body solver, and that it is also robust to adding massive neutrinos to the cold DM component. However, this discovery raises crucial questions about the validity of this invariance in MG models. Thus, we examined whether the invariance observed by Alimi \& Koskas remains robust in the case of Hu \& Sawicki model using \dustp numerical simulations. By comparing the results of advanced numerical simulations in these different theoretical frameworks, we found significant deviations from the invariance observed in the framework of $w$CDM models of GR. These deviations suggest that the gravitation's nature significantly influences the DM halos' shape. We then interpreted this departure from the GR models' invariance as a manifestation of the scalar-field screening effect corresponding to such $f(R)$-type theories. This one modifies the sphericization process of DM halos during their formation, precisely because the critical mass at which this scalar field becomes non-negligible is the mass at which the deviation appears. To this extent, the departure from cosmological invariance in DM halos' shape is a cosmological probe of the nature of gravity, and the mass scale at which it appears can be used to estimate the $f_{R0}$ parameter of such theories.
}
\maketitle
\nolinenumbers
\section{Introduction}
In their ambition to elucidate the enigmatic nature of dark energy, responsible for the observed cosmic acceleration, cosmology researchers have increasingly moved beyond the standard $\Lambda$CDM paradigm anchored in Einstein's theory of general relativity \citep[GR;][]{Lue:2003ky,
2012PhR...513....1C, Joyce2016, Langlois:2018dxi}. The emerging extended gravity models offer convincing alternatives, introducing pronounced deviations from GR that generate accelerated cosmic expansion in the later epochs of the Universe without a cosmological constant or an exotic energy component. These deviations not only influence homogeneous cosmic dynamics, but they also considerably modify the growth of density perturbations, leaving unmistakable imprints on the formation of cosmic structures \citep{song2007, Bertschinger:2008zb, Shant2010, bernardeau2011, koyama2015, 2018IJMPD..2748005H, Sharma:2022fiw}. Remarkably, in some alternative gravitational frameworks, the fundamental principles of GR can resurface at higher densities through screening mechanisms, ensuring compatibility with high-precision gravitational data derived from Solar System studies \citep{Khoury:2003aq, PhysRevD.76.063505, osti_20935321, brax2008, jain2010, Liu:2017xef, Sakstein:2018fwz}. Typical examples of these modified gravity (MG) theories are models of the $f(R)$ type \citep{Starobinsky:1980te, Capozziello2003CURVATUREQM, caroll2004}, in which the standard scalar curvature, $R$, in the Einstein-Hilbert action, is replaced by a more general function, $f(R)$. Within this new theoretical framework, the approach of \citet{Hu:2007nk}, hereafter referred to as HS, aims not only to harmonize with the homogeneous cosmic dynamics of the widely accepted $\Lambda$CDM model, but also to provide new elements of cosmological structuring through the introduction of an additional fifth force. \\ \\
It has also been obtained that gravity modifications affect differently the lensing potential and the dynamical potential, so that estimates of cluster mass using either photon deflection or the velocity of their massive constituents no longer match \citep{Schmidt:2010jr, 2016A&A...595A..78G}. The density profiles of clusters due to matter accumulation in their accretion region in HS models are also modified \citep{Lombriser:2011zw}. However, the search for a distinguishing feature to differentiate models built in the framework of GR from those built in the framework of modified gravity (MG) theories remains a crucial open question.\\ \\
\citet{Llinares} and \citet{Burrage:2014daa} suggested that the study of the shape of dark matter (DM) halos in MG models could be a more critical way of discriminating between the MG models and models developed in GR. Indeed, due to the screening process inherent to these models, the collapse during the formation of DM halos naturally leads to an anisotropic component that can accentuate the triaxiality of these cosmic structures. The work presented in this article addresses, from a purely statistical perspective, the crucial question of whether such a feature in the shape of DM halos could imprint a specific MG signature in structure formation processes. We investigated the influence of MG-$f(R)$ models on the statistical properties of the shape of DM halos. We also tested whether the cosmological invariance of the shape properties of DM halos, when expressed in terms of the nonlinear fluctuations of the cosmic matter, already verified for $w$CDM models of GR (as we had demonstrated in \citet{alko} from the Dark Energy Universe numerical simulations \citep{Alimi2010, Rasera2010aip, Alimi2012, Reverdy2015}), can be extended to HS models performed by \citet{Giocoli:2018gqh}.\\ \\
It is important to note that the cosmological invariance of the shape properties of DM halos resulted both from the isotropic nature of the gravitational interaction in $w$CDM models of GR, implying that the more a halo collapses, the more it evolves toward a spherical cosmic structure, and from the fact that the estimate of the nonlinear fluctuations also results from an isotropic calculation of the two-point correlation in the cosmic matter field. Ultimately, this cosmological invariance expressed that the shape of DM halos in $w$CDM models of GR encapsulates the entire power spectrum of the cosmic matter field \citep{alko}. It remains to be seen whether the anisotropy enhancement due to scale-dependent, and therefore mass-dependent, screening processes during the formation of DM halos in MG models would break, or at least somehow modify, this cosmological universality property on the shape of DM halos \citet{alko}.\\ \\
Our paper is structured as follows. We first present (Sect. \ref{methodology}) a comprehensive review and interpretation of the findings reported by \citet{alko} concerning $w$CDM models from RG and we discuss the potential physical limitations of such cosmological universality beyond those models. We then present the theoretical framework of the HS theories and \dustp simulations carried out by \citet{Giocoli:2018gqh}, detailing the models simulated and the DM halo catalogs derivable from such simulations. In Sect. \ref{triax}, we explore the shape properties of the DM halo populations in MG models in terms of mass and terms of the nonlinear fluctuations of the cosmic matter field. We discuss how MG models can break cosmological invariance observed on $w$CDM models in GR. We then estimate the scale at which universality could be broken analytically, precisely because of the screening effect. We then discuss that, although universality may be broken, the universal correspondence (according to cosmology) between the nonlinear fluctuations of matter within halos and the nonlinear fluctuations of all cosmic matter on the halo scale is preserved. In the final section, we summarize our main findings and their implications, and discuss future directions for this line of research. 
\section{Methodology}
\label{methodology}
\subsection{Motivations: Fundamental cosmological invariance of the shape properties of DM halos}
\label{motivations}
\begin{figure}[h]
\centering
\includegraphics[width=8.5cm]{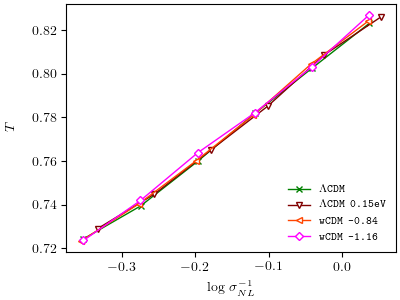}
\caption{Median triaxiality of the DM halos of the $w$CDM models from the \dustp simulations (with or without neutrinos), as function of the non-linear fluctuations of the cosmic matter field, $\sigma_{NL}$. All the curves are superimposed, which means that the cosmological dependence of the mass-triaxiality relations is once again completely encapsulated in the non-linear power spectrum, as was highlighted in \citet{alko}.}
\label{invariance1}
\end{figure}

\citet{alko} highlighted the existence of a fundamental cosmological invariance of the shape properties of DM halos in $w$CDM models from GR. Indeed, they first showed that shape parameters such as the prolaticity $p$, the triaxiality $T$, or the ellipticity $E$ of a DM halo strongly depended on the mass of the DM halo and that such dependence was moreover highly sensitive to the underlying cosmology. However, when these shape parameters were re-expressed in terms of the nonlinear fluctuations of cosmic matter, this cosmological dependence completely disappeared.\\ \\
The shape parameters $p, T, E$ are defined from the eigenvalues ${a ^2}/{5}$, ${b^2}/{5}$, ${c^2}/{5}$ of the total inertia tensor $\mathcal{M}^{\mathrm{FoF}}$: \begin{equation*}
  \mathcal{M}^{\mathrm{FoF}}_{1\leq i, j\leq 3}=\langle x_ix_j\rangle - \langle x_i\rangle \langle x_j\rangle\;, 
\end{equation*} where the averages are taken over all the particles included in the DM halo. The prolaticity of halos is \begin{equation*}p=\frac{a-2b+c}{2(a+b+c)}\;,\end{equation*} which is worth $0$ for spheres, $+\infty$ for cigar-shaped halos, and $-\infty$ for pancake-like halos, while the triaxiality of halos, \begin{equation*}T=\frac{a^2-b^2}{a^2-c^2}\;,\end{equation*} tends to be $1$ for cigar-shaped halos and $0$ for pancake-like halos. Finally, the ellipticity, \begin{equation*}E=\frac{a-c}{2(a+b+c)}\;,\end{equation*} quantifies the deviation from sphericity. The cosmological invariance has been highlighted from the DM halos catalog, built from the $w$CDM models' Dark Energy Universe numerical simulation data \citep{Alimi2010, Rasera2010aip, Alimi2012, Reverdy2015}. In principle, as shown in \citet{alko}, the calculation of the shape parameters $p, T, E$ of DM halos should be carried out with caution to avoid potential numerical and methodological biases. Indeed, such biases can arise from a variety of sources, such as limited numerical resolution, particle sampling effects, halo search algorithms, or assumptions made in shape measurement techniques \citet{Zemp2011}. To avoid such potential numerical effects, the set of particles taken into account when calculating the total inertia tensor must be carefully defined \citep{alko}. 
This is because, as the numerical resolution in mass increases, the substructures inside the halos are better resolved, and because they are generally off-center, they dominate the inertia tensor of the halo in which they are located. The shape of the halo is then artificially dominated by the shape of the substructures on its edge and, therefore, depends on the resolution. This effect can be very significant and can, if not corrected, distort the whole analysis. The correction process is explained in detail in \citet{alko}. However, in this article, all the simulations used (see below) have the same resolution (and the same number of particles); consequently, the only remaining precaution to take when measuring the shapes of the DM halos is to consider their ellipsoidal fit, deduced from the diagonalization of the total inertia tensor of all the particles in the halo detected by the FOF algorithm \citep{alko}, which we did. \\ \\
\citet{Giocoli:2018gqh} performed simulations for models of the same nature (where only the equation-of-state parameter $w$ is modified) as those used in \citet{alko}. Using the output data, we computed median triaxiality $T$ (or prolaticity $p$, or ellipticity $E$) of DM halos of these models\footnote{All results presented in this paper for median quantities are also recovered for mean quantities.}, detected by the FOF algorithm. In Fig. \ref{invariance1}, we plot $T$ as function of the nonlinear fluctuations of the total cosmic matter field, smoothed over the mass scale of these halos. The $w$CDM models of \dustp simulations are represented, including models with and without neutrinos. All curves superpose and show the invariance we highlighted in \citet{alko}; this means that cosmological dependence of mass-shape relations is again completely encapsulated in the nonlinear power spectrum. We note that the presence of neutrinos does nothing to change the existence of this cosmological invariance. The results are analogous when we plot the ellipticity or prolaticity of the DM halos. 
It was also shown in \citet{alko} that cosmological invariance is verified either if the shape of the DM halos is expressed in terms of nonlinear fluctuations of the "full" cosmic matter field (smoothed on the halo scale) or in terms of nonlinear fluctuations of the cosmic matter field only within the DM halos. Again, this result is also very well verified here. When we re-express the shape parameters in terms of the nonlinear fluctuations of the cosmic matter only inside the halos, they appear invariant in the same way according to the cosmology. This latter property then induces the existence of a correspondence between the nonlinear fluctuations of the matter inside the DM halos and those of the total cosmic field as in \citet{alko}; once again, this is insensitive to the underlying cosmology (see Sect. \ref{mapping}).\\ \\
These first results obtained with the \dustp simulations \citep{Giocoli:2018gqh} also show that the invariance highlighted in \citet{alko} is robust when the numerical method is changed; that is, a different N-body solver is used: the Dark Energy Universe simulations were carried out using an optimized version of the RAMSES code \citep{Teyssier, Alimi2012}, which is an AMR multigrid code. Conversely, the \dustp simulations were carried out using an advanced version of the GADGET Tree code \citep{springel2005, puchwein}. \\ \\
In GR, under the effect of gravitation, the forming DM halos tend to become more and more spherical over
time. However, along their formation, their shape properties stay independent of cosmology when expressed in terms of the nonlinear fluctuations of the cosmic field. Our goal is to elucidate the fate of this invariance when the gravitational collapse is modified as expected in a theory of MG. Indeed, the sphericization process during the gravitational collapse would be altered, because screening effects can influence the collapse differently along each axis during DM halo triaxial formation.
We rely on the DM halos of HS cosmological models, as simulated by \citet{Giocoli:2018gqh}.
\subsection{Hu \& Sawicki models in MG theories}
\label{hu}
\begin{figure}
\centering
\includegraphics[width=8.5cm]{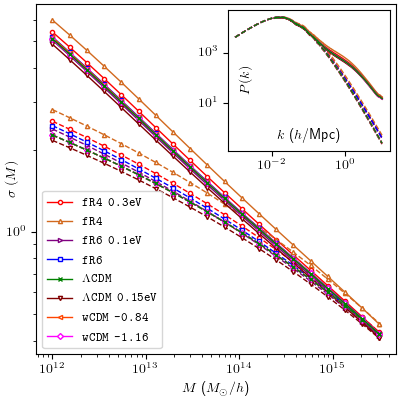}
\caption{Root-mean-square of nonlinear fluctuations of cosmic matter field as function of the smoothing mass M. The linear fluctuations $\sigma_L$ (computed with MGCAMB) are represented by dashed lines, while the solid curves correspond to the $\sigma_{NL}$, the nonlinear fluctuations of cosmic matter field (computed from the $z = 0$ snapshots of the \dustp simulations). The internal panel contains the corresponding linear and nonlinear power spectra.}
\label{sigma-P}
\end{figure}

In $f(R)$ cosmologies, the behavior of matter is governed by revised Einstein field equations. A direct adaptation of GR that bypasses the cosmological constant, $\Lambda$, is achieved by incorporating a function of the curvature scalar into the Einstein-Hilbert action. As a result, a curvature-dependent term appears in the equations of motion, mimicking the effects of dark energy. In these models, the modified Einstein-Hilbert action can be written as
\begin{equation}
\mathcal{L}_{HS}=\frac{c^3}{16\pi G}\int \mathrm{d}^4x\sqrt{-\lvert g\rvert}\left(R+f(R)\right),\; 
\end{equation}
\noindent where $g$ denotes the metric tensor, $G$ is Newton's gravitational constant, $R$ is the curvature scalar, and $f(R)$ is a scalar function responsible for an additional degree of freedom. For a comprehensive review of GR modifications, we advise consulting
\citet{2012PhR...513....1C}. \\ \\
To ensure consistency between the different scales, HS considers a specific $f$ function compatible with the strict Solar System constraints while accurately representing the cosmic dynamical expansion of the $\Lambda$CDM model. In the HS model, this function takes the following form:
\begin{equation}
f(R)=-m^2\frac{c_1\left(\frac{R}{m^2}\right)^n}{c_2\left(\frac{R}{m^2}\right)^n+1}\;.
\end{equation}
\noindent Here, $n$, $c_1$, and $c_2$ are the model parameters. $m^2 \equiv \Omega_m H_0^2$, where $\Omega_m$ is the current matter density parameter and $H_0$ is the present Hubble parameter. In this paper, we consider $n=1$. \citet{Hu:2007nk} showed that the constraints imposed by the cosmic background radiation and the late expansion of the Universe force the function $f(R)$ to be:
\begin{equation}
f(R)={-6(1-\Omega_{m})\cdot R}\cdot\left(\frac{R}{H_0^2}+\frac{\lvert f_{R0}\rvert\left(12-9\Omega_{m}\right)^2}{6(1-\Omega_{m})}\right)^{-1}.
\end{equation}
\noindent Therefore, compared to GR, the sole supplementary degree of freedom is $f_{R0}$. It can be seen in the low curvature limit as the present-day characteristic amplitude of the dimensionless scalar field $f_R\equiv \mathrm{d}f/\mathrm{d}R$. The standard model is recovered when $\lvert f_{R0}\rvert$ tends toward $0$.\\ \\
From the revised Einstein field equations, one obtains the revised Friedmann equations \citep{song2007}. When this HS $f(R)$ theory is rewritten in the form of a massive scalar-tensor model with potential, an important aspect of such a model emerges: the mass of the dynamical scalar field varies with the local density \citep{Hu:2007nk}. This so-called chameleon mechanism \citep{Khoury:2003aq, Khoury:2003rn, Brax:2004qh} leads to the disappearance of MG effects in high-density regions so that gravity then behaves as GR. We cover this fundamental property and its consequences in more detail later.\\ \\
\citet{Giocoli:2018gqh} performed the \dustp simulation suite. These N-body simulations simulate a volume of $(750\, \textrm{Mpc}/h)^3$ with $N_p = 768^3$ particles (equivalent to a mass resolution of $m_p \approx 8\cdot 10^{10} M/h$). The \dustp simulation suite contains flat $w$CDM models (including $\Lambda$CDM concordance model) and HS-$f(R)$ models, optionally including massive neutrinos. We now focus on the simulations where $f_{R0}$ takes extreme values: successively $10^{-4}$ (\texttt{fR4}; high value) and $10^{-6}$ (\texttt{fR6}; low value). The other standard cosmological parameters are consistent with the Planck-2015 cosmological data: $\Omega_m = 0.31345$, $\Omega_b = 0.0481$, $H_0 = 67.31$, km/s/\textrm{Mpc}, $n_s = 0.9658$, and $\sigma_8 = 0.847$. These simulations were carried out using the MG-gadget simulation code \citep{puchwein}. For an exhaustive overview of these simulations, we refer the reader to \citet{Giocoli:2018gqh, Hagstotz:2018onp, Hagstotz:2019gsv, Lee:2022lbu}.\\ \\
First, we discuss the impact of deviations from GR on the formation of cosmic structures through the properties of the \dustp power spectra.\\ \\
Using MGCAMB \citep{camb, Zhao:2008bn, Hojjati:2011ix, Zucca:2019xhg, Wang:2023tjj}, we computed the linear spectra of the cosmic matter field for these models at $z=0$, denoted $P_L$, and the variance of these linear power spectra:
\begin{equation}
\label{eq4}
\sigma^2_L(M)=\int_{0}^{+\infty}k^2 P_L(k) W^2\left[k\cdot \left(\frac{3M}{4\pi\Omega_m\rho_c}\right)^{1/3}\right]\;\frac{\mathrm{d}k}{2\pi^2}\;, 
\end{equation}
\noindent where $W$ is a Gaussian window function: $W(x)=\exp -\frac{x^2}{10}$. In fact, $\sigma^2_L(M)$ measures the level of the collapse of the cosmic field on mass scale $M$ compared to the average mass one expects for a homogeneous field ($4\pi \Omega_m\rho_c R^3/3$). 
$P_L$ and $\sigma_L$ are represented by dotted lines in Fig. \ref{sigma-P}. We observe that increasing $f_{R0}$ increases the variance and hence the clustering in the Universe. This is particularly true when we focus on the low-mass range.\\ \\
We proceeded in the same way for the nonlinear power spectrum, $P_{NL}(k)$; $P_{NL}$ is calculated 
from the particle positions of the last snapshot ($z=0$) of the simulations, by applying a CIC interpolation scheme to a $3500^3$ grid of voxels. The $P_{NL}$s are represented by solid lines in the inner panel of Fig. \ref{sigma-P}. We note that the nonlinear dynamics induces much stronger clustering than the linear dynamics; indeed, $P_{NL}^{\mathrm{\Lambda CDM}}(k=1)$ is four times higher than $P_{L}^{\mathrm{\Lambda CDM}}(k=1)$, and, similarly, $\sigma_{NL}^{\mathrm{\Lambda CDM}}(M=10^{14})\approx 1.2 \sigma_{L}^{\mathrm{\Lambda CDM}}(M=10^{14})$. For large masses ($M>10^{15} M_\odot/h$) or large scales ($k<0.2 \, h/\textrm{Mpc}$), the linear and nonlinear regimes merge as expected. 
In the \dustp simulations suite, the initial power spectra are the same in all models. The ratio $P_L/P_L^{\mathrm{\Lambda CDM}}$ then gives the growth rate of each model relatively to the one of the $\Lambda$CDM model. These growth rates are scale-dependent in MG models, and their main feature is the increase in the gravitational constant ($G_{\mathrm{eff}}=4G/3$) at the smallest spatial scales. The collapse of low-mass halos is therefore enhanced when the amplitude of the scalar field greatly exceeds that of the Newtonian potential. It should also be noted that, when neutrinos are included, they tend to homogenize the distribution of cosmic matter and therefore attenuate this accentuation of the gravitational interaction. For instance, the linear spectrum of \texttt{fR4} is 60\% larger than in the concordance model, whereas \texttt{fR4-0.3eV} only shows an increase of 30\% (at $k=1 \; h/\textrm{Mpc}$). 
\begin{figure}
\centering
\includegraphics[width=8.5cm]{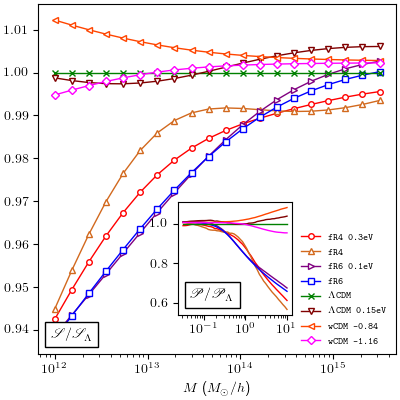}
\caption{Purely nonlinear variance relatively to $\Lambda_{CDM}$ one as function of the smoothing mass $M$. For a given model of gravity, the purely nonlinear variance is defined as $\mathscr{S}={\sigma_{NL}}/{\sigma_L}$. In the internal panel, we similarly plot, for each model, the purely nonlinear power spectrum $\mathscr{P}={P_{NL}}/{P_L}$ relatively to the $\Lambda${CDM} one.}
\label{ratio-sigma}
\end{figure}
However, such linear enhancements combine with the effects of nonlinear dynamics. To distinguish between the contributions of the linear dynamics and those of the nonlinear dynamics due specifically to the change in gravity, we compute, for every model, the pure, nonlinear power spectrum $\mathscr{P}={P_{NL}}/{P_L}$ and, similarly, the pure, nonlinear variance $\mathscr{S}={\sigma_{NL}}/{\sigma_L}$. The ratio of these last quantities to their value for $\Lambda$CDM model, the pure nonlinear boost, is plotted in Fig. \ref{ratio-sigma} for each cosmological model. This gives, for each $f_{R0}$, the gain in power brought by nonlinear dynamics compared to that obtained in the concordance model. We observe that, for the MG models, the purely nonlinear spectrum quickly decreases and is $20$\% weaker than that of $\Lambda$CDM as early as $k=1\, h/\textrm{Mpc}$. At the smallest scales or masses, the collapse of cosmic matter generates high-density environments, as the mass of the dynamic scalar field is then very high - this leads it to reach the minimum of the potential in which it evolves. Effects of chameleon-field existence therefore vanish and gravity mimics the Newtonian gravitation ($G_{\mathrm{eff}}=G$); the nonlinear power spectrum no longer carries the accentuated effect of MG, whereas, at the same scale, the linear spectrum keeps doing so. As a result, the screening effect mitigates the pure, nonlinear boost at high $k$ (small scale or small mass). This would explain why the nonlinear boost decreases sharply at large $k$ and why such a decrease is initiated on a scale that depends on the chosen $f_{R0}$.
As the \dustp simulations are also designed to study the degeneracy between gravity modification and neutrino mass, we refer the reader to \citet{Giocoli:2018gqh} for further details on their joint effects on the power spectrum.
\begin{table*}[h!]
\centering
\caption{\small Parameters of \dustp simulations and the main features of the associated halo catalogs.}
\label{params}
\begin{tabular}{|c||c|c|c|c|c|c|c|c|c|c|}
\hline
Name 
& $\lvert f_{R0}\rvert$ 
& $\sigma_8^{\mathrm{Gauss}}$ 
& $m_\nu$ 
& $\Omega_m$ 
& $w_0$ 
& $N$ 
& $M_{FoF}^{\mathrm{Minimal}}$ 
& $M_{FoF}^{\mathrm{Median}}$ 
& $M_{FoF}^{\mathrm{Mean}}$ 
& $M_{FoF}^{\mathrm{Maximal}}$ 
\\ 
\hline
\hline
\texttt{$\Lambda$CDM} & 0 & $0.88$ & $0$ & $0.31345$ & $-1$ & $270,256$ & $8.1$ & $16$ & $33.6$ & $5520$ 
\\ 
\hline
\texttt{$\Lambda$CDM 0.15eV} & 0 & $0.85$ & $0.15$ & $0.30987$ & $-1$ & $266,159$ & $8.01$ & $15.7$ & $32.3$ & $5090$
\\ 
\hline
\texttt{wCDM -0.84} & 0 & $0.88$ & $0$ & $0.31345$ & $-0.84$ & $273,209$ & $8.1$ & $16$ & $33.7$ & $5580$
\\ 
\hline
\texttt{wCDM -1.16} & 0 & $0.88$ & $0$ & $0.31345$ & $-1.16$ & $268,307$ & $8.1$ & $16$ & $33.6$ & $5480$
\\ 
\hline
\texttt{fR6} & $10^{-6}$ & $0.89$ & $0$ & $0.31345$ & $-1$ & $299,555$ & $8.1$ & $15.7$ & $32.6$ & $5530$
\\ 
\hline
\texttt{fR6 0.1eV} & $10^{-6}$ & $0.87$ & $0.1$ & $0.31107$ &
$-1$ & $296,127$ & $8.04$ & $15.5$ & $31.8$ & $5250$
\\ 
\hline
\texttt{fR4} & $10^{-4}$ & $1.0$ & $0$ & $0.31345$ 
&$-1$ & $289,815$ & $8.1$ & $16.7$ & $37.9$ & $6200$
\\ 
\hline
\texttt{fR4 0.3eV} & $10^{-4}$ & $0.93$ & $0.3$ & $0.3063$ &$-1$ & $285,912$ & $7.92$ & $15.9$ & $34.3$ & $3310$
\\ 
\hline
\end{tabular}
\tablefoot{
Masses are in units of $10^{12}$ $(M_\odot/h)$ and neutrino mass $m_\nu$ is in eV. The quantity $\sigma_8$ is computed using the linear power spectrum and a Gaussian smoothing window (see Eq. \ref{eq4}). 
Finally, $N$ is the number of halos. }
\end{table*}
\subsection{DM halo datasets of the \dustp simulations}
\label{dustgrain}
Beyond the analysis of the power spectrum of the cosmic density field, we now study the shape of the halos and in particular how the cosmological invariance highlighted in the $w$CDM models from GR \citep{alko} can be altered by a non-vanishing $f_{R0}$; this enables us to go further in our understanding of the impact of the chameleon screening.\\ \\
To do so, we built halo catalogs from \dustp simulations with or without massive neutrinos. The halos were identified using the friends-of-friends (FOF) algorithm, taking the percolation parameter $b=0.2$. We chose to stay consistent with the convention adopted in \citet{Giocoli:2018gqh}, in which massive neutrinos were ignored during the FoF detection. This is justified in \citet{castorina2014} by the fact that the very large free streaming of massive neutrinos can spuriously link them to CDM particle groups.
\\ \\\Cref{params} features these catalogs' properties for every simulation used in this article. From these new numerical datasets, we tested the possible extension of the cosmological invariance of the distribution of the DM halos shape in terms of the nonlinear fluctuations of the cosmic matter field, particularly for Hu \& Sawicki MG models. 
\section{Triaxiality of DM halos in MG models}
\label{triax}
\subsection{Halo shape measurement}
\begin{figure}
\centering
\includegraphics[width=8.5cm]
{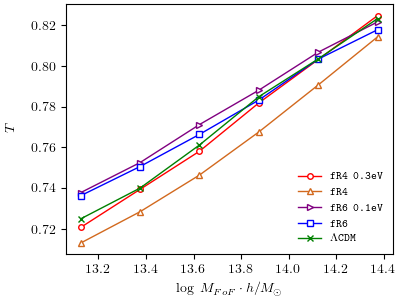}
\caption{Median triaxiality of DM halos as function of their FoF mass. Each curve corresponds to a given gravitational and cosmological model, with or without massive neutrinos. The most massive halos are the most prolate, and the higher the $\sigma_8$, the more oblate the halos.}
\label{TM}
\end{figure}

\begin{figure}
\centering
\includegraphics[width=8.5cm]{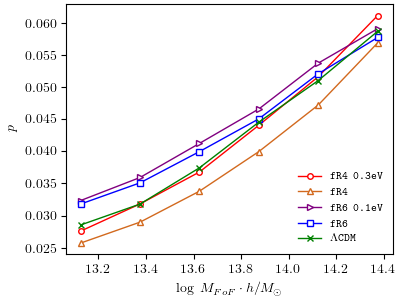}
\caption{Median prolaticity of DM halos as function of their FoF mass.}
\label{PM}
\end{figure}

\begin{figure}
\centering
\includegraphics[width=8.5cm]{"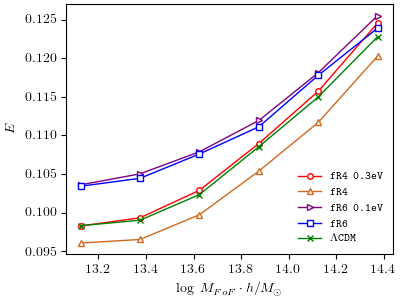"}
\caption{Median ellipticity of DM halos as function of their FoF mass. Again, DM halo ellipticity increases with mass and decreases with $\sigma_8$. The DM halos become more and more spherical as they collapse.}
\label{EM}
\end{figure}

As all the numerical simulations examined in this article share the same resolution (the same number of particles and box size), it is sufficient to incorporate all the particles associated with a given FOF halo in the mass tensor (whose eigenvalues give an ellipsoidal fit of the considered halo). The precautions taken to evaluate the shape of the halos concerning substructure removal would have been necessary if our simulations had been of different resolutions \citep{alko}, but this is not required here.\\ \\
For the mass interval $\left[10^{13}, 10^{14. 5 }\right]$ $M_\odot/h$, in Figs. \ref{TM}, \ref{PM}, and \ref{EM}, we plot the median values of the shape parameters (respectively, $T$, $P,$ and $E$) as function of their mass. In the same way as for models of the $w$CDM type, from Dark Energy Universe simulations \citep{alko} or from simulations of the \dustp suite (see Sect. \ref{motivations}), these shape parameters increase with the mass of the DM halo. The lighter the DM halo, the more spherical and flattened it tends to be. The uncertainties in the measurements of the shape parameters of DM halos are tiny, and, consequently, the error bars that could be added to the figures in this article are almost indistinguishable from the line thickness. Indeed, two types of error bars can be distinguished; the first is statistical. It is fixed by the number of halos from which the median of the shape parameters is calculated. This statistical error is negligible as the number of halos is very large in each catalog (> 250,000; see Table \ref{params}) in each mass bin. A second kind of uncertainty is systematic. It is fixed by the number of particles on which the inertia tensor (and therefore the shape parameters) of each DM halo are calculated. The selected DM halos contain several hundred particles, so this error is also tiny (cf. Appendix B in \citet{alko}). Therefore, possible uncertainties in shape measurements do not affect the results presented in this article. \\ \\
The dependence of the shape properties on the mass of the DM halos is a consequence of the radial and isotropic nature of Newtonian gravitational collapse. In the mass domain considered here, the triaxiality increases by about $13\%$ (from 0.73 to 0.82), the prolaticity doubles (i.e., from 0.030 to 0.060), and the ellipticity increases by $25\%$. Figures \ref{TM}, \ref{PM}, and \ref{EM} also show that, at a fixed mass, $E$, $p,$ and $T$ decrease as the $\sigma_8$ parameter of the current cosmological model increases. The higher $\sigma_8$, the more the structures are collapsed (at a given redshift) and, consequently, the more spherical they are. The corner piece of gravity modification effects on the halo shapes lies in the fact that the chameleon screening is effective at a given $f_{R0}$-dependent mass, around which it can alter the gravitational collapse of DM halos. It indeed introduces an anisotropic component in the collapse. To quantify the impact of the latter, we study the shape dependence on cosmic matter field nonlinear fluctuations (rather than on the mass) and compare it to the dependence given by a Newtonian collapse.
\subsection{Breaking of the cosmological invariance of DM halos shape in MG models.}
\label{mapping}
\begin{figure}[h]
\centering
\includegraphics[width=8.5cm]{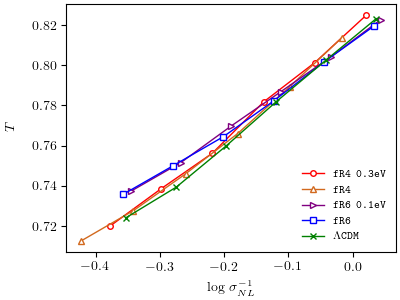}
\caption{Median triaxiality of DM halos as function of nonlinear fluctuations of cosmic matter field, $\sigma_{NL}$. All the curves superpose, which means that the cosmological dependence of the mass-triaxiality relations is completely encapsulated in the nonlinear power spectrum. However, when $\lvert f_{R0}\rvert=10^{-6}$, whatever the neutrino mass, the relation between $\sigma_{NL}$ and triaxiality does not identify with the relation obtained for the other models. This occurs specifically from $\log(\sigma_{NL}^{-1})\approx-0.1$, corresponding to a mass of $10^{14}\, h/M_\odot$ for the
\texttt{fR6} model, which is, for this model, the characteristic mass where the screening starts. The way the halos collapse and sphericize therefore depends on the gravity model. The screening phenomenon modifies the evolution of the shape of the DM halos, which can no longer be encapsulated in the power spectrum. The divergence in the $\sigma_{NL}-T$ relations is then a distinctive marker of the change in the gravity model, which nevertheless remains independent of the neutrino mass. }
\label{invariance}
\end{figure}

\begin{figure}[h]
\centering
\includegraphics[width=8.5cm]{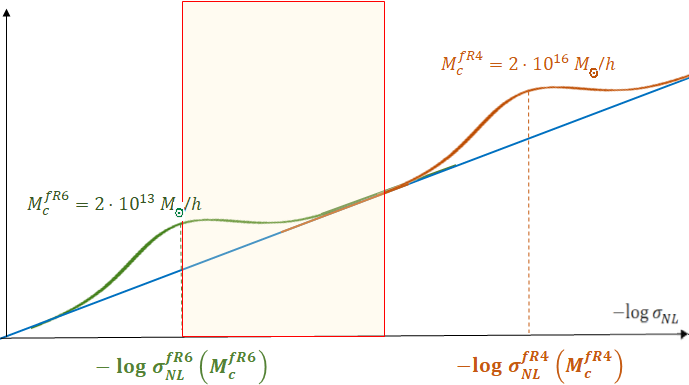}
\caption{Sphericization process of DM halos during their formation, according to their mass and the screening effect of MG models. The mass range of the DM halos detected in the numerical simulation is delimited by the red rectangle. The a-sphericization process affects DM halos around the critical mass depending on the present-day characteristic amplitude of the dimensionless scalar field $f_{R0}$ of each model (see Eq.  (7)). These DM halos are indeed present for the \texttt{fR6} numerical simulation, but not the \texttt{fR4} numerical simulation.}
\label{invarianceG}
\end{figure}

\begin{figure}[h]
\centering
\includegraphics[width=8.5cm]{"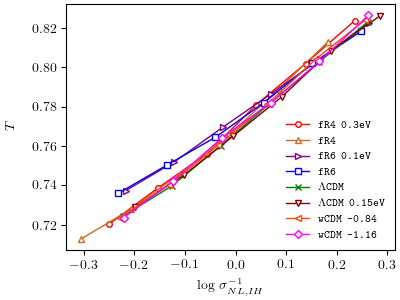"}
\caption{Median triaxiality of halos as function of variance of non-linear fluctuations of cosmic matter internal to DM halos. All the curves superpose, meaning again that the cosmological dependence of mass-triaxiality relations is completely encapsulated in the nonlinear power spectrum. However, when $\lvert f_{R0}\rvert=10^{-6}$, independently of the neutrino mass, the median relation between variance and triaxiality does not identify with the relation obtained for the other models.}
\label{fig11}
\end{figure}

\begin{figure}[h]
\centering
\includegraphics[width=8.5cm]{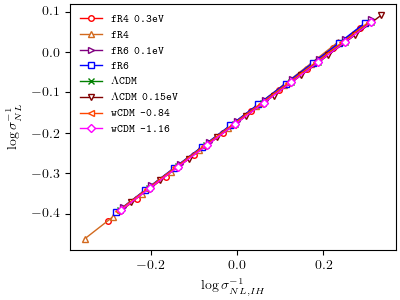}
\caption{Scatter plot of $\left(\sigma_{NL, IH}, \sigma_{NL}\right)$ for every value of $f_{R0}$ and $M_{\nu}$. There is manifestly a correspondence between those two kinds of nonlinear variance, respectively computed from the matter inside the halos and from the whole cosmic matter field. this correspondence seems to be universal, independently of the cosmological parameters and the gravity model.}
\label{fig12}
\end{figure}

We now consider cosmological models that vary the gravity parameter ($f_{R0}$) as well as the potentially non-vanishing neutrino mass ($m_\nu$). We want to know if the cosmological invariance of the shape parameters as function of the nonlinear fluctuations of the cosmic matter field highlighted in \citet{alko}, and again found for the $w$CDM models of the \dustp suite (Sect. \ref{motivations}), disappears for such models. In Fig. \ref{invariance}, we observe that replacing $M$ by $\sigma_{NL}(M)$ again compensates for the cosmological dependence of the triaxiality of the halos: all the curves merge, regardless of the specific cosmological and gravitational parameters, but there is one notable exception. While the \texttt{fR6} curves, with or without massive neutrinos, indeed superpose well on the other curves at large masses (i.e., at large values of $\log\sigma_{NL}^{-1}$), they differ from the others if $\log\sigma_{NL}^{-1}<-0.15$. This corresponds for these two models to a typical mass of $10^{14}$ $M_\odot/h$. Prolaticity and ellipticity measurements show similar behaviors. \\ \\
It seems that the origin of this invariance breaking can be backtracked to the halo collapse process. Indeed, since the gravitational isopotentials of a given ellipsoidal distribution have the remarkable property of being systematically more spherical than the isodensities that give rise to them \citep{Llinares, Stapelberg2020}, the halos become more and more spherical as their collapse progresses. Meanwhile, as the halos collapse, the cosmic matter field becomes more structured, and its nonlinear fluctuations increase, so that there necessarily exists a relationship (independent of cosmology) between the triaxiality (or prolaticity or ellipticity) of halos of mass $M$ on the one hand,  and the nonlinear fluctuations of the cosmic matter field measured here with $\sigma_{NL}(M)$ on the other hand. Adding a screening mechanism will change the picture since it is more or less effective, depending on the potential well depth and (therefore) the halo mass; it will consequently lead to a change in the sphericization process of the halos.\\ \\
We distinguish three mass regimes for the screening mechanism. High-mass halos, whose absolute Newtonian potential ${GM}/(Rc^2)$
is much larger than $\lvert f_{R0}\rvert$, are completely screened. Gravity is then indistinguishable from Newtonian gravitation. Such halos become more and more spherical with time, and no significant difference in their shape is expected compared to the halos accreted in a purely Newtonian context. The relation between $T$ and $\sigma_{NL}$ for these halos is then unchanged compared to the one we obtained for the $w$CDM halos (which assume GR). On the contrary, very low-mass halos, whose potential is low compared to $\lvert f_{R0}\rvert $, were not (or very slightly) screened. Gravity behaves again like Newtonian gravitation, with a modified effective gravitational constant of $G_{eff}\simeq{4}G/{3}$. As such a modification of gravity remains perfectly isotropic, the invariance of the shape properties as function of nonlinear fluctuations of the cosmic field is again preserved; In reality, halos sphericize faster, but the cosmic matter field clusters faster too. \\ \\
In these two cases, the observed cosmological dependence shape according to the mass is then fully absorbed into the power spectrum of the cosmic matter field in which the DM halos formed. In other words, for these halos, the relationship between $\sigma_{NL}$ and $T$ remains identical to that observed in the $w$CDM models from GR. \\ \\
The situation is more subtle when the (Newtonian) potential of the halos is comparable, in terms of absolute value, to $f_{R0}$. 
As \citet{Llinares} pointed out, these halos can be screened along a single principal axis of their ellipsoid and not be screened along the other axes. In MG, such halos can then become more elliptical along the axis where the screening mechanism was effective. In other words, as the chameleon field is generally as elliptical as the isodensities, its presence tends to alter the efficiency of the "sphericization" expected in Newtonian gravity; the ellipsoidal matter distribution of such halos, therefore, tends to sphericize over time less than is expected in purely Newtonian gravity. At constant density, the critical potential (of the order of $f_{R0}$) can be recast as a critical mass ($M_c$). The triaxiality, the prolaticity, or the ellipticity of the DM halos whose $\sigma_{NL}$ is close to $\sigma_{NL}(M_c)$ will be higher in MG than for DM halos of the same $\sigma_{NL}$ in the $w$CDM models.\\ \\
To find an explicit expression of $M_c$, one can use the fact that a DM halo is screened as soon as \citep{Baker:2019gxo}
\begin{equation}
\label{eqsc}
\frac{GM}{Rc^2}\geq\frac{3}{2}\lvert f_{R0}\rvert\;.
\end{equation}
\noindent As the average spherical density of FoF halos (with $b=0.2$) is generally around 200$\rho_m$, which gives \begin{equation*}\frac{M}{R}\geq M/\left(\frac{\frac{3}{4\pi} M}{200\rho_m}\right)^{1/3},\end{equation*}
\Cref{eqsc} becomes 
\begin{equation*}
GM^{2/3}\geq\frac{3c^2}{2}\lvert f_{R0}\rvert \left(\frac{3}{800\pi\rho_m}\right)^{1/3}\;.
\end{equation*}
\noindent Hence, the typical mass of the screened halos is 
\begin{equation}
M_c=\left(\frac{3c^2}{2G}\lvert f_{R0}\rvert\right)^{3/2}\left(\frac{3}{800\pi\rho_m}\right)^{1/2}\;.
\end{equation}
\noindent Assuming $\Omega_m=0.31$, we obtain 
\begin{equation}
M_c=\lvert f_{R0}\rvert^{3/2}\left(2.1\cdot 10^{22} M_\odot/h\right).
\end{equation}\\ \\
For the \texttt{fR4} model, $M_c=2.1\cdot 10^{16}M_\odot/h$. For such a model, all the DM halos have a mass well below the critical mass at which the screening effect becomes effective. The halos are then said to have low masses, and the ($\sigma_{NL}-T$) curve of \texttt{fR4} merges with that of the models derived from GR (Fig. \ref{invarianceG}). For the \texttt{fR6} model, on the other hand, since $M_c=2.1\cdot 10^{13}M_\odot/h$, the DM halos' masses are comparable to this critical mass, and they are only likely to be screened during their formation along some of their principal axes \citep{Llinares}. The sphericization process is then altered, and the $\sigma_{NL}-T$ invariance is broken around such a critical mass as that observed in Fig. \ref{invariance} (i.e., for halos with a mass lower than about $10^{14} M_\odot/h)$. Above this mass, for more massive halos that are completely screened, the invariance is again well verified (Fig. \ref{invariance}). The modifications of the sphericization process due to the screening effect, according to the value of $f_{R0}$ and the mass of the DM halos, are summed up in the diagram of Fig. \ref{invarianceG}. \\ \\
\Cref{invariance} shows the relation between triaxiality and nonlinear fluctuations of the cosmic matter field for multiple values of $f_{R0}$ and $m_\nu$. The results for prolaticity and ellipticity flow in the same direction. Although our interpretation is confirmed by these results, it cannot be considered sufficient. Indeed, the combined effect of gravitational collapse and the shielding effect due to the chameleon scalar field in modified-gravity models, and how it modifies the sphericization process of the forming halo, is obviously very complex. This combined effect may evolve and even possibly reverse during the formation of the halo, favoring either an a-sphericization or a sphericization. To improve the simplistic and binary nature of our interpretation, we would need a larger number of high-resolution numerical simulations with a wide range of well-sampled $f_{R0}$ values and a wider range of DM halo masses, screened or not. Such simulations are not yet available, so we defer such a detailed analysis to future work. However, we believe that the interpretation we are proposing, although binary, is a good first proposal compatible with the given data. 
As observed in Fig. \ref{invariance}, it is worth noticing that the presence of massive neutrinos does not change the previous conclusions: the $\sigma_{NL}-T$ curves of the same gravity models, with or without massive neutrinos, are systematically merged. \\ \\
The $\sigma_{NL}$-shaped curves thus become a probe of MG at the limit of the mass domain where the screening effect is initiated. The joint measurement of the nonlinear power spectrum of a given cosmic matter field, on the one hand, and of the shape of the halos it contains, on the other, will then make it possible to interpret any break in the curve, known to be universal for the classical gravity cosmological models,  as the manifestation of a change in gravity. The scale of the nonlinear fluctuations where the invariance break occurs also provides an estimate of $f_{R0}$. \\ \\
If the shape of the halos is, by definition, calculated on the halos themselves (practically meaning from the particles that make up the halos) and independently of their environment, it would then be logical to only compute the nonlinear fluctuations on the matter inside the halos, rather than the total cosmic matter field. Let $\sigma_{NL, IH}$ denote the nonlinear fluctuations of the cosmic matter inside the halos: 
\begin{equation*}
\sigma^2_{NL, IH}(M)=\int_{\mathbb{R}_+}k^2 P_{NL, IH}(k) W^2\left[k\cdot \left(\frac{3M}{4\pi\Omega_m\rho_c}\right)^{\frac{1}{3}}\right]\;\frac{\mathrm{d}k}{2\pi^2}\;, 
\end{equation*} 
\noindent where the power spectrum $P_{NL, IH}(k)$ is computed by taking into account only the particles belonging to the halos. 
In Fig. \ref{fig11}, we plot the median triaxiality of the halos in bins of $\log\sigma_{NL, IH}^{-1}$ . The results are quite similar to those obtained with $\log\sigma_{NL}^{-1}$ - the curves merge for all cosmological models (Fig. \ref{fig11}), whereas the invariance is again broken at low masses for \texttt{fR6} halos (Fig. \ref{fig11}). \\ \\
The fact that, at high mass, there is a universal relationship between $T$ and $\sigma_{NL}$ on the one hand and between $T$ and $\sigma_{NL, IH}$ on the other, must imply a correspondence (independent of cosmology and the gravity model) between $\sigma_{NL}$ and $\sigma_{NL, IH}$ in the domain corresponding to high mass, similarly to the correspondence highlighted in \citet{alko}. To verify this and to find out what happens at lower masses, in Fig. \ref{fig12} we plot $\sigma_{NL}(M)$ in terms of $\sigma_{NL, IH}(M)$ for each simulation and the entire mass domain covered by the simulation halos (i.e., $[10^{13}, 10^{14.5}]\;M_\odot/h$). We find that over the whole mass domain, even when including $M_c$, the correspondence between $\sigma_{NL}$ and $\sigma_{NL, IH}$ is well verified, independently of the cosmology and gravity model. 
\section{Discussion and conclusion}
In this work, we investigated the effect of MG on the shape of DM halos in the special case of the Hu-Sawicki modified-gravity models. We show that the variance of the nonlinear power spectrum, smoothed over the scale of the halos, completely determines the triaxiality of the halos whatever the cosmological model, except the light halos (mass less than $10^{14}\, M_\odot/h$) for the $f_{R0}=-10^{-6}$ models from \dustp simulation data. We propose the following interpretation of this result: the higher the nonlinear variance smoothed on the scale of the halos, the more structured the corresponding Universe is and the more its halos are collapsed. Under the effect of gravitation, the halos become spherical as they collapse. This confirms the existence of a relationship between the degree of structuring of the Universe (a measure of which is given by the variance of the power spectrum) and the triaxiality of the halos it contains. However, this interpretation of a universal relationship independent of the cosmology, between the power spectrum variance and the triaxiality of DM halos, is broken if we compare different models of MG (and therefore different sphericization processes). The universality relationship linking nonlinear fluctuations of cosmic matter field and shapes of DM halos must finally be understood "for a given gravity model". In the specific context of Hu-Sawicki models, it results from the phenomenon of screening inherent in these models, with a critical mass dependent on $\rho_m$ and especially the parameter controlling the deviation from GR $\lvert f_{R0}\rvert$. DM halos whose mass is then close to this critical mass are likely to be screened (by a chameleon mechanism) along one of their principal axes only so that the total gravity field is much more elliptical than the Newtonian isopotentials. The sphericization process is then altered, while the power spectrum (and therefore its variance), which is averaged over the angles, ignores the information linked to the induced anisotropies of this type. Thus, for a given parameter $\lvert f_{R0}\rvert$, there is still a relationship between the nonlinear fluctuations of cosmic matter field and triaxiality (or any other shape parameter) of the DM halos, which does not depend on the other cosmological parameters (neutrino mass, $\Omega_M$, $\sigma_8$, $w_0$...). However, this relationship differs from that found in pure $w$CDM models in the vicinity of this critical mass, and the simultaneous measurement of the power spectrum and shape of the halos thus becomes an innovative probe of MG. \\ \\
It then becomes exciting to study the effects of other MG models on the shape of the halos, with different screening mechanisms. Examples include symmetron screening and nDGP screening. In addition, any change including the nature of DM could also, in principle, be likely to leave a visible imprint on the shape of the halos. The formation of hot DM halos or fuzzy halos differs greatly from those observed in the case of cold DM. This could lead to the relationship between the shape of the halos and the power spectrum being given a fundamental nature, with regard to which any deviation could be seen as a powerful probe of a change in gravity or the nature of DM. 
 \begin{acknowledgements} 
 We would like to thank Dr Philippe Brax for fruitful discussions.
 \end{acknowledgements}
 
\bibliographystyle{aa}
\bibliography{alimiandkoskas}
\appendix
\section{Broken degeneracy on $\delta_c$}
\begin{figure}[h]
\centering
\includegraphics[width=8.5cm]{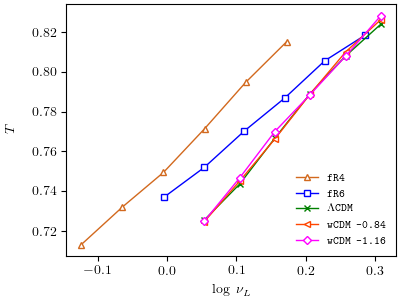}
\caption{Triaxiality of DM halos at $z=0$ as function of the variable $\nu_L(M, 0)=\delta_c(z=0)/\sigma_L(M, 0)$ which takes into account the explicit dependence on the critical spherical overdensities $\delta_c$, of each model. We do not observe any full cosmological invariance.} \label{fig9}
\end{figure}

\begin{figure}[h]
\centering
\includegraphics[width=8.5cm]{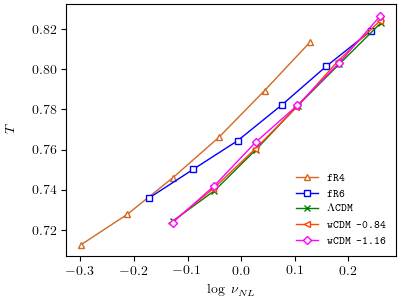}
\caption{Triaxiality of DM halos at $z=0$ as function of the variable $\nu_{NL}(M, 0)=\delta_c(z=0)/\sigma_{NL}(M, 0)$ which takes into account the explicit dependence on the critical spherical overdensities $\delta_c$, of each model. We note the universality for the cosmological models with the same $\delta_c(z=0)$ parameter, but We observe no full cosmological invariance. } 
\label{fig10}
\end{figure}

The models we study in this paper admit very different critical spherical overdensities $\delta_c$ at $z=0$ (see Table \ref{params2}). It is therefore necessary to test whether an invariance in terms of the variables $\nu_L(M, z)=\delta_c(z=0)/\sigma_L(M, z)$ or $\nu_{NL}(M, z)=\delta_c(z=0)/\sigma_L(M, z)$ that combine both the cosmic matter field fluctuations and the $\delta_c$ effect exists.
\begin{table}[h!]
\centering
\caption{Critical spherical overdensities $\delta_c$ at $z=0$ for cosmological models with or without MG.}
\label{params2}
\begin{tabular}{|c||c|}
\hline
Name & $\delta_c(M=10^{14}\, M_\odot/h)$\\ 
\hline
\hline
\texttt{$\Lambda$CDM} & $1.68$\\ 
\hline
\texttt{wCDM -0.84}&$1.68$\\ 
\hline
\texttt{wCDM -1.16} &$1.68$\\ 
\hline
\texttt{fR6}&$1.59$\\ 
\hline
\texttt{fR4} & $1.38$\\ 
\hline
\end{tabular}
\end{table}
We plot respectively in Figs. \ref{fig9} and \ref{fig10} the median triaxiality of DM halos as function of $\nu_L$ and $\nu_{NL}$. We see clearly that the cosmological invariance is not recovered. There is no invariance in terms of $\nu_L$ and we observe no rapprochement of the curves when we consider $\nu_{NL}$. It is uniquely the re-expression of the shape properties of the DM halos in terms of the non-linear fluctuations of the cosmic matter field that allows the highlighting of a cosmological invariance for a given theory of gravity. The inclusion of massive neutrinos has no bearing on this question.
\\ \\
\end{document}